# Принцип соответствия и эволюция физики

## Ю.И. Богданов

*Физико-технологический институт РАН* [1]
*Москва, Нахимовский пр.,34*

**Аннотация**

На основе принципа соответствия Н. Бора показано, что релятивистская механика, как и квантовая механика могут рассматриваться как рациональные обобщения механики классической. Сравнительное изложение релятивистской и классической механик проведено с использованием трех основных положений: определения импульса, основного закона динамики (второго закона Ньютона) и закона сохранения энергии. Отличие релятивистской механики от классической обусловлено новым определением импульса как меры количества движения, пропорциональной скорости и энергии. Изучение связи между энергией и импульсом приводит к получению импульсно- энергетических инвариантов для классических и релятивистских систем. Эти инварианты имеют важное значение при описании взаимодействий. Показано, что новые релятивистские законы динамики приводят к необходимости видоизменения кинематических соотношений классической механики, таких как закон сложения скоростей, преобразование координат и др. Показано, что квантовая механика может рассматриваться как рациональное статистическое обобщение классической механики. Статистические закономерности в квантовой механике носят фундаментальный объективный характер и не связаны с неполнотой информации об изучаемой системе. Среди возможных многопараметрических статистических моделей выделенную роль играет корневая модель, связанная с введением амплитуд вероятностей (пси функции) как математического объекта статистического анализа данных. Построение многопараметрической статистической модели сводится к нахождению таких частот и базисных функций в разложении Фурье, которые обеспечивали бы выполнение в среднем классических уравнений движения. Корневая модель приводит к согласованному условию, связывающему собственные частоты и функции механической системы и выражаемому матричным уравнением Гейзенберга. Матричное уравнение Гейзенберга сводится к операторному уравнению, решение которого можно интерпретировать как построение гамильтониана системы и переход к картине Шредингера. Рассматриваемый подход естественным образом приводит к понятию оператора импульса, фундаментальным коммутационным соотношениям, построению матрицы плотности, уравнения Лиувилля и др.

Работа носит методический характер. Рассчитана на студентов, преподавателей и научных работников, интересующихся методологическими вопросами становления современной физики.

---

[1] e-mail: bogdanov@ftian.oivta.ru

**Введение**

По решению Организации Объединенных Наций, текущий- 2005 год объявлен всемирным годом физики. Весьма примечательно, что этот год является юбилейным для двух, быть может, самых важных открытий в физике XX века. Речь идет о релятивистской механике, которой исполняется 100 лет и квантовой механике, которой исполняется 80 лет.

Возникновение релятивистской механики связывают с написанными в 1905 году двумя работами А. Пуанкаре и двумя работами А. Эйнштейна. Обе работы Пуанкаре имеют одно общее название «О динамике электрона» [1], причем вторая является расширенной версией первой. Первая из указанных работ доложена Пуанкаре 05 июня 1905 года в Академии наук в Париже (в том же месяце работа вышла и докладах Академии). Вторая работа Пуанкаре поступила в редакцию 23 июля 1905 года и вышла в 1906 году. Первая работа А. Эйнштейна «К электродинамике движущихся тел» [2] поступила в редакцию 30 июня 1905 года, вторая- «Зависит ли инерция тела от содержащейся в нем энергии» [3] - 27 сентября 1905 года. Необходимо отметить, что указанные выше основополагающие работы обоих авторов, наверное, никогда бы не появились на свет, если бы не было многолетней деятельности Лоренца, направленной на объединение электродинамики Максвелла с электронной теорией (электродинамика движущихся тел занимала очень важное место в этой деятельности). Таким образом, Лоренц должен считаться создателем релятивистской механики в не меньшей степени, чем Пуанкаре или Эйнштейн. Важно также подчеркнуть крупный вклад Минковского, представления которого о четырехмерном пространстве- времени сделали всю теорию, в известной мере, завершенной. Таким образом, с исторической точки зрения, релятивистская механика должна называться теорией Лоренца- Пуанкаре- Эйнштейна- Минковского.

Второй, 80-летний квантовый юбилей, связан с работой В. Гейзенберга «О квантовотеоретической интерпретации кинематических и механических соотношений» [ 4] (поступила в редакцию 29 июля 1925 года), а также с двумя последовавшими за ней работами под общим названием «К квантовой механике» [5-6] (первая из этих работ написана М. Борном и П. Иорданом, а вторая совместно В. Гейзенбергом, М. Борном и П. Иорданом). В указанных работах заложены основы так называемой матричной механики. Они знаменует собой начало «новой» квантовой теории, пришедшей на смену «старой», развивавшейся в период с 1900 г. (когда Планк ввел в физику свой знаменитый квант действия). Отметим, что на протяжении многих лет, как до 1925 года, так и после, решающую роль в становлении идеологии квантовой теории играл Н. Бор.

Настоящая работа посвящена представленным выше двум теориям-юбилярам. Нашей целью будет показать, как две новые механики (релятивистская и квантовая) исторически и логически выросли из старой (классической) механики. Мы предприняли попытку, по возможности, элементарного, но строгого изложения физических основ релятивистской механики и квантовой теории с точки зрения их преемственности по отношению к классической механике. Надеемся, что представленный материал, в первую очередь, будет полезен студентам как введение в предмет и задачи современной физики. С методической точки зрения, данная работа, как попытка заполнения «бреши» между современной и классической физикой, возможно, привлечет внимание также преподавателей и научных работников.

В основу изложения положен **принцип соответствия** Н. Бора. Согласно этому принципу, современная физика не отрицает классическую, а выступает как ее рациональное обобщение, содержащее классическую теорию в качестве частного (предельного) случая.

Теория относительности (релятивистская механика), ровно как и квантовая механика, есть своего рода «аналитические» продолжения (рациональные обобщения) классической ньютоновской механики. Эти «аналитические продолжения» связаны с введением в теорию двух новых фундаментальных констант: предельной скорости движения и распространения взаимодействий в природе $c$, равной скорости света в вакууме, а также кванта действия $\hbar$.

Символические формулы, выражающие принцип соответствия, хорошо известны. Это требование, чтобы релятивистская механика переходила в классическую механику при устремлении скорости света к бесконечности ($c \to \infty$), т.е. когда все, характерные для рассматриваемой задачи скорости, малы по сравнению с предельно возможной скоростью, а также аналогичное требование, чтобы квантовая механика переходила в классическую механику при $\hbar \to 0$ (т.е. в задачах, в которых несущественна конечная величина кванта действия). Последний переход достаточно нетривиален, т.к. при $\hbar \to 0$ возникает сингулярность. Тем не менее, формализм этого перехода хорошо разработан (так называемое квазиклассическое приближение в квантовой механике).

Приведенные выше утверждения пока только декларируют принцип соответствия, но не раскрывают еще его содержания. Содержательная часть этого принципа раскрывается только при конкретном изложении взаимоотношений «новых» механик (релятивистской и квантовой) со «старой» классической механикой. Заметим здесь только, что вопреки выводу, который может возникнуть при поверхностном рассмотрении предмета, принцип соответствия совсем нетривиален. Он, в частности, не сводится к одному только методическому и философскому пожеланию, чтобы новая теория не отрицала начисто старую, а только ограничивала бы сферу ее действия. Содержание принципа соответствия глубже. Достаточно сказать, что Н. Бор выдвинул принцип соответствия в 1918 г. до создания последовательной квантовой теории. «Стремление рассматривать квантовую теорию как рациональное обобщение классических теорий привело к установлению так называемого принципа соответствия» ([7], с.41). Принцип соответствия Бора служил физикам в то время, по выражению Зоммерфельда, «волшебной палочкой» для получения с помощью классической теории нетривиальных результатов, относящихся к еще неизведанной тогда области квантовых явлений ([8], с.193). Другими словами, принцип соответствия может служить для того, чтобы «заглянуть» за границу классической теории с помощью самой классической теории (что, конечно, нетривиально). Граница старой теории, в этом смысле, не есть глухая стена, а, скорее, это- борт корабля, отделяющий освоенную территорию, от внешней стихии, на которую можно взглянуть с борта корабля, но в которую лучше не погружаться без специального снаряжения (в виде акваланга или батискафа, несущих частицу корабля).

Заслуга классической механики в том, что уже в ее рамках были введены такие фундаментальные понятия как импульс и энергия, а также понятие взаимодействия (как источника перераспределения импульсов и энергий составляющих систему частиц). Универсальность понятий импульса и энергии обусловлена их органической связью с однородностью пространства и времени [9, 10]. В этом причина исключительной «живучести» фундаментальных классических понятий. Классическая и современная физика построены, в сущности, на едином концептуальном базисе. Можно сказать, что современная физика- все еще ньютоновская. Возможно, что когда- нибудь возникнет другая (неньютоновская) физика, которая будет построена на совершенно другом концептуальном базисе. Это означает, что принцип соответствия верифицируем (то есть может, в принципе, быть отвергнут дальнейшим развитием науки). Мы не утверждаем, что этот принцип имеет всеобщий характер, а констатируем только, что развитие физики до сих пор проходило в соответствии с этим принципом.

С точки зрения принципа соответствия, провозглашенная Бором необходимость использования классической механики для описания работы экспериментальной установки в атомной физике, является вполне логичной и не должна рассматриваться как недостаток квантовой механики. Соответствие между квантовой механикой и классической, в какой- то мере, аналогично соответствию между комплексными и действительными числами. Понятие комплексного числа, очевидно, является рациональным обобщением понятия действительного числа. Однако, комплексное число невозможно описать иначе, как пару действительных чисел. Точно также, и результаты экспериментов в квантовой механике невозможно описать иначе, как посредством классических экспериментальных установок. Более того, точно также как комплексное число есть совокупность двух (так сказать, взаимно- дополнительных) действительных чисел, полное описание квантового объекта задается не иначе как совокупностью различных взаимно- дополнительных классических наблюдений.

Согласно принципу дополнительности Н. Бора «данные, получаемые при разных условиях опыта, не могут быть охвачены одной- единственной картиной; эти данные должны скорее рассматриваться как *дополнительные* в том смысле, что только совокупность разных явлений может дать более полное представление о свойствах объекта» [11].

Требовать, чтобы квантовая механика была описана независимо от классической- это все равно, что требовать, чтобы теория комплексных чисел была бы сформулирована независимо от понятия действительного числа.

## Часть 1. Две системы механики: классическая и релятивистская

> Nature and nature's laws lay hid in night.
> God sad: "Let Newton be!" And all was light.
>
> Александр Поп (1688-1744), цитируется по [12], с.212.

### 1.1. Основные положения классической и релятивистской механики

В качестве атрибутов, характеризующих движение материальной частицы в механике, выступают три основные величины: **импульс**, определяющий меру (количество) движения тела, **сила**, определяющая взаимодействие как причину изменения импульса, а также **энергия**, изменение которой определяется работой сил, действующих на частицу. Примечательно, что всё отличие релятивистской механики от классической определяется только различием в определении импульса.

В классической механике Ньютона импульс определяется формулой:

$$\vec{p} = m \cdot \vec{v} \qquad (1.1)$$

«Количество движения (*импульс*) есть мера такового, устанавливаемая пропорционально скорости и массе» (И. Ньютон «Математические начала натуральной философии», цитируется по [13], с.73)

В релятивистской механике это определение модифицируется таким образом, что коэффициентом пропорциональности между импульсом и скоростью оказывается энергия, а не масса.

$$\vec{p} = \frac{E}{c^2} \cdot \vec{v} \qquad (1.2)$$

Перефразируя слова И. Ньютона, уточненное релятивистской физикой определение импульса можно сформулировать так: «Количество движения (*импульс*) есть мера такового, устанавливаемая пропорционально скорости и энергии».

Определение импульса в релятивистской механике должно рассматриваться как постулат, который вводится взамен классического определения (которое тоже постулат). Значение и содержание нового постулата выявляются, конечно, только апостериори (в результате развития теории и сравнения новых предсказаний теории с экспериментом). Использование энергии вместо массы приводит к упрощению и унификации теории. Вместо двух законов сохранения в классической механике- закона сохранения (или принципа аддитивности) масс и закона сохранения механической энергии (который может нарушаться действием неконсервативных сил), мы имеем в новой (релятивистской) механике один всеобщий закон сохранения энергии. Таким образом, «закон сохранения энергии, поглотив ранее закон сохранения тепла, включил теперь в себя и принцип сохранения массы и управляет всем единолично» ([14] с.655).

Чтобы параметр $E/c^2$ имел размерность массы, нужно чтобы параметр $c$ имел размерность скорости. Развитие механики на основе нового определения импульса приводит к экспериментально подтверждаемому выводу о том, что $c$ - максимально возможная скорость движения в природе (см. разд. 1.3,1.4), совпадающая со скоростью света в вакууме.

Основной закон динамики (**второй закон Ньютона**) справедлив в обеих теориях и имеет один и тот же вид:

$$\Delta \vec{p} = \vec{F} \cdot \Delta t \qquad (1.3)$$

Этот постулат утверждает, что изменение импульса тела (левая часть равенства (1.3)) обусловлено импульсом силы (правая часть (1.3)).

В дифференциальной форме основной закон динамики можно записать так:

$$\frac{d}{dt}\vec{p} = \vec{F} \qquad (1.4)$$

Заметим, что второй закон Ньютона обычно (например, в школе) записывают в форме $\vec{F} = m\vec{a}$, которая перестает быть верной при переходе в релятивистскую область (здесь $\vec{a}$ - ускорение). Замечательно, что свой основной закон динамики сам Ньютон записал сразу в форме (1.3) – (1.4), которая остается корректной и в релятивистской механике. Это обстоятельство, на наш взгляд, не совсем случайно. Очевидно, Ньютон хорошо понимал, что взаимодействие, по своей внутренней природе, является причиной изменения импульса, а изменение скорости (ускорение) выступает при этом уже как вторичное явление.

В качестве третьего основного положения механики выберем **закон сохранения энергии.** Изменение энергии тела определяется работой действующих на него сил. Соответствующий энергетический баланс определяется уравнением:

$$\frac{dE}{dt} = \vec{v} \cdot \vec{F} \qquad (1.5)$$

Уравнение (1.5) показывает, что скорость изменения энергии есть работа в единицу времени, то есть мощность, равная скалярному произведению силы на скорость. Выбор энергии в качестве одного из основных понятий механики выглядит наиболее естественно в релятивистской механике. Действительно, здесь энергия введена изначально - как коэффициент пропорциональности между скоростью и импульсом. Для того, чтобы сделать теорию внутренне замкнутой, вполне естественно потребовать в релятивистской механике сохранения энергии, которое, таким образом, выступает вместо классического принципа сохранения массы и, кроме того, как будет видно ниже, вполне заменяет термодинамическое понятие внутренней энергии.

Релятивистская механика, в определенном смысле, ставит точку в длинном историческом споре между сторонниками Декарта (1596- 1650) и Лейбница (1646-1716) о том, какая из величин (импульс или кинетическая энергия) служит более естественной мерой движения. Ньютон вслед за Декартом в качестве меры движения выбирает импульс (импульс у Декарта, правда, еще является скалярной величиной). В 1686 г., однако, Лейбниц публикует статью, в которой критикует импульс Декарта в качестве меры движения и предлагает в качестве сохраняющегося в природе «количества двигательной активности» свою «живую силу» (удвоенную кинетическую энергию в современных обозначениях). В это время (1686 г.) «Математические начала натуральной философии» Ньютона уже были написаны, но еще не были изданы («Начала» вышли из печати только в следующем 1687 году).

"Живая сила" Лейбница, однако, не могла стать универсальной характеристикой движения. Кинетическая энергия классической теории, в силу незамкнутости самой классической механики по отношению к неупругим (тепловым и т.п.) процессам, не «дотягивает» еще до универсальной меры, способной характеризовать как «внешнее», так и «внутреннее» движение тела. Этому критерию вполне удовлетворяет только релятивистское понятие энергии. Таким образом, релятивистское обобщение классической механики делает ее основы более простыми и логически замкнутыми. Новая механика показывает, что импульс и энергия неразрывно связаны между собой уже в силу определения (1.2).

Рассмотрим теперь вытекающую из основных положений теории связь между энергией и импульсом.

### 1.2. Связь между энергией и импульсом в механике Ньютона.

Используя закон сохранения энергии (1.5), определение импульса (1.1) и основной закон динамики (1.4), получаем очевидную цепочку тождеств:

$$\frac{dE}{dt} = \vec{v} \cdot \vec{F} = \frac{\vec{p}}{m} \frac{d\vec{p}}{dt} = \frac{d}{dt} \frac{\vec{p}^2}{2m}, \qquad (2.1)$$

откуда имеем следующую сохраняющуюся величину:

$$\frac{d}{dt}\left(E - \frac{\vec{p}^2}{2m}\right) = 0 \qquad (2.2)$$

Учитывая, что $\vec{p}^2 = p^2$, получаем для энергии тела выражение:

$$E = \frac{p^2}{2m} + E_0 = T + E_0 \qquad (2.3)$$

где $E_0$ - внутренняя энергия тела, т. е. энергия, отвечающая случаю, когда тело в целом покоится ( $\vec{p} = 0$ ), хотя его составные части, конечно, могут двигаться. Кинетическая энергия определяется выражением:

$$T = E - E_0 = \frac{p^2}{2m} = \frac{mv^2}{2} \qquad (2.4)$$

**1.3. Связь между энергией и импульсом в релятивистской механике**

В полной аналогии с проведенными выше выкладками, получим теперь выражение для связи релятивистских энергии и импульса. Используя закон сохранения энергии (1.5), определение релятивистского импульса (1.2) и основной закон динамики (1.4), получаем:

$$\frac{dE}{dt} = \vec{v} \cdot \vec{F} = \frac{\vec{p}c^2}{E} \frac{d\vec{p}}{dt}, \qquad (3.1)$$

откуда $E \dfrac{dE}{dt} = c^2 \vec{p} \dfrac{d\vec{p}}{dt}$, (3.2)

поэтому $\dfrac{d}{dt}\left(E^2 - p^2 c^2\right) = 0$, (3.3)

Таким образом для связи релятивистских энергии и импульса получаем выражение:

$$E^2 - p^2 c^2 = E_0^2 \qquad (3.4),$$

где $E_0$ - внутренняя энергия тела.

Энергия $E$ и импульс $\vec{p}$ частицы, конечно, зависят от выбора системы отсчета. Однако, как показывает соотношение (3.4), если из квадрата энергии вычесть квадрат импульса, помноженный на квадрат скорости света, то получим уже величину, которая не зависит от выбора системы отсчета, т.е. является релятивистским инвариантом (этот инвариант есть квадрат внутренней энергии).
Учитывая выражение (1.2) для релятивистского импульса, получим:

$$E^2 \left(1 - \frac{v^2}{c^2}\right) = E_0^2, \qquad (3.5)$$

откуда имеем для зависимости энергии от скорости выражение:

$$E = \frac{E_0}{\sqrt{1 - \frac{v^2}{c^2}}} \qquad (3.6)$$

Из выражения (3.6) видно, что $E \to \infty$ при $v \to c$. Уже отсюда очевиден предельный характер скорости света (см. также пример, рассмотренный в разделе 1.4). Рассмотрим нерелятивистское приближение для формулы (3.6). Если $v \ll c$, то, разложив (3.6) в ряд по малому параметру $v/c$, и, ограничиваясь первым неисчезающим слагаемым, получим:

$$E = E_0 + \frac{E_0}{c^2} \frac{v^2}{2} \qquad (3.7)$$

В силу принципа соответствия**,** при малых скоростях (по сравнению со скоростью света) релятивистская механика должна переходить в классическую, т.е. выражения (2.3) и (3.7) должны тождественно совпадать.

Таким образом, из принципа соответствия, получаем знаменитую формулу, дающую связь между внутренней энергией тела и его массой (эту формулу, как правило, называют формулой Эйнштейна).

$$E_0 = mc^2 \qquad (3.8)$$

Из полученного результата следует, что при изменении внутренней энергии тела, меняется и его масса (нагретое тело, в принципе, тяжелее холодного, существует очень слабый эффект изменение массы при сгорании химического горючего, весьма заметный дефект масс в ядерных превращениях и т.д.)

Очень большая по житейским меркам величина скорости света не давала возможности (скажем, в период между открытием закона сохранения энергии в термодинамике и созданием релятивистской механики) для осознания и экспериментального обнаружения тождества между внутренней энергией системы и ее массой. Недопонимание этого фундаментального свойства приводит и до сих пор к путанице, по крайней мере, терминологической. Так, вместо правильного и естественного, как с исторической, так и логической точек зрения, термина «внутренняя энергия», до сих пор общеупотребительным остается явно неадекватный и устаревший термин «энергия покоя» (см., например, классический учебник [15]). Термин «энергия покоя» возник, очевидно, как производное понятие от устаревшего термина «масса покоя» (массу частицы в правой части (3.8) было принято называть массой покоя). Развернутая критика термина «масса покоя» дана в работе Л.Б. Окуня [16]. Автор [16], однако, справедливо критикуя термин «масса покоя» как явно устаревший, оставляет без внимания ничуть не лучший термин «энергия покоя».

Введение лишнего термина «энергия покоя» неизбежно породило иллюзию, будто «энергия покоя» и «внутренняя энергия»- это различные понятия. Такая иллюзия, возможно, могла существовать некоторое время, пока связь между массой и внутренней энергией была продекламирована, но еще не была подтверждена, в достаточной степени, экспериментально в ядерной физике, а затем и в физике элементарных частиц. В наше время, повсеместное использование этого термина представляется явным анахронизмом.

Заметим, наконец, что термины «энергия покоя» и «масса покоя», в некотором смысле, «близнецы- братья». Использование термина «энергия покоя» неизбежно порождает своего двойника – «массу покоя».

Самая знаменитая формула физики $E_0 = mc^2$ раскрывает фундаментальную связь между массой тела и его колоссальной внутренней энергией. Подобно тому, как формула Джоуля 1кал=4,1868Дж задает механический эквивалент для теплоты, формула Эйнштейна $E_0 = mc^2$ задает энергетический эквивалент для массы.

Физическое содержание понятия массы заключается в том, что масса тела есть мера содержащейся в нем внутренней энергии. Перефразируя известное изречение Ф. Бэкона: «Теплота есть (внутреннее) движение и ничего более», можно сказать, что «масса есть внутренняя энергия и ничего более». Масса внутренне присуща телу, как и внутренняя энергия. Открытие их эквивалентности есть такое же великое достижение, как и, скажем, открытие Фарадеем и Максвеллом единства электрических, магнитных и световых явлений.

Так называемое (еще один неудачный термин) превращение «массы в энергию» и обратное превращение «энергии в массу» есть не что иное, как превращение внутренней энергии исходной частицы (например, радиоактивного ядра) в кинетическую энергию осколков и энергию квантов полей (фотонов), а также обратное превращение энергии движения во внутреннюю энергию продуктов реакции при неупругих столкновениях.

Учитывая (3.8), получим зависящие от скорости выражения для энергии, и импульса в релятивистской механике.

$$E = \frac{mc^2}{\sqrt{1 - \frac{v^2}{c^2}}} \quad (3.9)$$

$$\vec{p} = \frac{m \cdot \vec{v}}{\sqrt{1 - \frac{v^2}{c^2}}} \quad (3.10)$$

С использованием массы связь (3.4) между энергией и импульсом можно представить в виде:

$$E^2 = p^2 c^2 + m^2 c^4 \quad (3.11)$$

Заметим, что существуют частицы с нулевой массой ($m = 0$). Это – хорошо известные фотоны (кванты электромагнитного поля), глюоны (кванты ядерного (сильного) взаимодействия) и, возможно, нейтрино, играющие важную роль в так называемом слабом взаимодействии (правда, в последние годы появляются новые весомые аргументы в пользу ненулевой массы нейтрино).

Для частиц с нулевой массой формулы (3.9) и (3.10) не используются, а формула (3.11) принимает вид

$$E = pc \quad (3.12)$$

Ничего мистического в таких частицах, конечно, нет: просто у них отсутствует внутренняя энергия, и вся их энергия связана с их движением (импульсом). При уменьшении импульса до нуля у частицы нулевой массы, исчезает и ее энергия, т.е. частица перестает существовать.

Кинетическая энергия в релятивистской механике, по определению, есть

$$T = E - mc^2 \quad (3.13)$$

Для массивных частиц в нерелятивистском пределе выражение (3.13) совпадает с определением кинетической энергии в классической механике.

Рассмотрим простой пример на применение релятивистских формул.

### 1.4. Движение вдоль прямой под действием постоянной силы

Пусть на частицу, первоначально находившуюся в состоянии покоя, в момент времени $t = 0$ начинает действовать постоянная сила $F = const$, направленная вдоль некоторой прямой.

Тогда основной закон динамики даст:

$$\frac{d}{dt}\frac{mv}{\sqrt{1 - \frac{v^2}{c^2}}} = F, \qquad (4.1)$$

откуда, сразу получаем:

$$\frac{v}{\sqrt{1 - \frac{v^2}{c^2}}} = \frac{F \cdot t}{m} \qquad (4.2).$$

Выразим из (4.2) скорость как функцию времени:

$$\frac{v}{c} = \frac{F \cdot t}{m \cdot c \sqrt{1 + \left(\frac{Ft}{mc}\right)^2}} \qquad (4.3)$$

Полученный результат легко проанализировать. На начальном участке движения, когда $Ft \ll mc$, имеет место нерелятивистское равноускоренное движение. В противоположном (ультрарелятивистском) приближение, когда $Ft \gg mc$, скорость тела с течением времени асимптотически стремится к скорости света (оставаясь, конечно, всегда ниже последней). Рассматриваемый пример схематично описывает процесс накопления энергии заряженной частицей в современных ускорителях (роль силы при этом играет высокочастотное электромагнитное поле).

### 1.5. Система частиц.

Тело, которое мы выше рассматривали как единое целое, на самом деле может состоять из некоторых частиц (включая и кванты полей).
Суммарная энергия и импульс системы представляют собой сумму вкладов от отдельных частиц:

$$E = E_1 + E_2 + ... + E_n \qquad (5.1)$$

$$\vec{p} = \vec{p}_1 + \vec{p}_2 + ... + \vec{p}_n \qquad (5.2)$$

Соотношение (1.8) запишется в виде

$$E^2 - p^2 c^2 = \left(E_1 + E_2 + ... + E_n\right)^2 - c^2 \left(\vec{p}_1 + \vec{p}_2 + ... + \vec{p}_n\right)^2 =$$
$$= E_0^2 = const \qquad (5.3)$$

Полученное выражение (5.3) задает импульсно-энергетический инвариант в релятивистской механике. Энергия и импульсы частиц имеют различные значения в зависимости от выбора системы координат, однако, если из квадрата суммарной энергии вычесть помноженный на квадрат скорости света суммарный импульс в квадрате, то всегда, независимо от выбора системы координат, будем получать одно и тоже число – константу, равную квадрату внутренней энергии.

Импульсно-энергетическому инварианту (5.3) можно придать геометрический смысл. Действительно, введем понятие четырех-импульса системы, три первые компоненты которого есть обычный импульс системы $p_{1,2,3} = \vec{p}$, а четвертая компонента связана с энергией системы $p_4 = iE/c$

$$p = (p_1, p_2, p_3, p_4) = \left( \vec{p}, \frac{iE}{c} \right) \qquad (5.4)$$

Тогда, тождество (5.3) с геометрической точки зрения можно будет интерпретировать как инвариантность квадрата четырех-импульса (скалярного произведение самого на себя) относительно вращений в четырехмерном псевдоевклидовом пространстве:

$$\sum_{\mu=1}^{4} p_\mu^2 = \vec{p}^2 - \frac{E^2}{c^2} = -\frac{E_0^2}{c^2} = inv \qquad (5.5)$$

Импульсно-энергетический инвариант в классической механике получается на основе выражения (2.3) и свойства аддитивности масс:

$$E - \frac{p^2}{2m} = (E_1 + ... + E_n) - \frac{(\vec{p}_1 + ... + \vec{p}_n)^2}{2(m_1 + ... + m_n)} = E_0 = const \qquad (5.6)$$

Существование импульсно-энергетического инварианта в классической механике означает, что хотя энергия и импульсы частиц имеют различные значения в зависимости от выбора системы координат, однако, если из суммарной энергии вычесть квадрат суммарного импульса, деленный на удвоенную суммарную массу, то всегда, независимо от выбора системы координат, будем получать одно и тоже число – константу, равную внутренней энергии системы.

Системы, описанные выше, очевидно, являются системами независимых (невзаимодействующих) частиц. Тем не менее, несмотря на то, что взаимодействие при таком подходе явно не учитывается, полученные выше результаты, допускают очень широкое использование, в частности, в так называемых столкновительных задачах атомной и ядерной физики. Здесь импульсно-энергетический инвариант позволяет связать между собой входящее (in) и выходящее (out) состояния. Сами же in- и out-состояния, как раз, и представляют собой системы невзаимодействующих частиц (in-система отвечает еще не взаимодействующим частицам, out-система отвечает уже не взаимодействующим частицам- продуктам реакции).

Рассмотрим два примера, описывающих применение соответственно нерелятивистского и релятивистского импульсно-энергетического инвариантов.

Пример 1. Найти пороговую (т.е. минимально возможную) кинетическую энергию $T$, которой должна обладать частица массой $m$ для возбуждения атома мишени ($Q$ - энергия возбуждения атома, $M$ - его масса)

Решение. Пусть $\varepsilon_0$ - внутренняя энергия частицы массой $m$, $E_0$ - внутренняя энергия исходного атома массой $M$. В лабораторной системе координат (в которой мишень и, следовательно, исходный атом первоначально покоились) весь импульс системы равен импульсу $\vec{p}$ налетающей частицы. Полная энергия системы в лабораторных координатах есть:
$$E_0 + \varepsilon_0 + \frac{p^2}{2m} \qquad (5.7)$$

Фундаментальный импульсно- энергетический инвариант в нерелятивистской теории есть полная энергия минус квадрат полного импульса, деленный на удвоенную полную массу. Инвариантность этой величины, отвечающая порогу реакции, задается условием:

$$E_0 + \varepsilon_0 + \frac{p^2}{2m} - \frac{p^2}{2(m+M)} = \varepsilon_0 + E_0 + Q \qquad (5.8)$$

Здесь в левой части записан импульсно- энергетический инвариант в лабораторной системе для in- состояния, а в правой – он же, но уже для out- состояния и в системе центра инерции (в которой, по определению, суммарный импульс равен нулю). Мы учли также, что порог отвечает условию, когда продукты реакции не имеют кинетической энергии в системе центра энергии, т.е. в этой системе имеем на выходе покоящуюся частицу с энергией $\varepsilon_0$ и покоящийся возбужденный атом с энергией $E_0 + Q$.

Из последнего равенства легко находим для пороговой энергии налетающей частицы:
$$T = \frac{p^2}{2m} = Q\frac{M+m}{M} \qquad (5.9)$$

Из полученного результата видно, что, например, при возбуждении атома водорода электроном ($m \ll M$) порог приблизительно равен $Q$, а при возбуждении протоном ($m \approx M$) соответственно $2Q$.

Не вошедшие в окончательный результат внутренние энергии $\varepsilon_0$ и $E_0$ сталкивающихся частиц, конечно, есть $\varepsilon_0 = mc^2$ и $E_0 = Mc^2$, однако в классической теории это неизвестно и несущественно.

Пример 2. Найти пороговую энергию в следующей реакции рождения протон- антипротонной пары при взаимодействии налетающего протона с покоящейся водородной мишенью ($m$ - масса протона).

$$p + p \to p + p + p + \widetilde{p} \qquad (5.10)$$

Здесь $p$ - протон, $\widetilde{p}$ - антипротон (частица той же массой, что и протон, но имеющая, в частности, противоположный протону электрический заряд).

Решение совершенно аналогично Примеру 1, но теперь необходимо воспользоваться релятивистскими формулами. Пусть $E$ - энергия налетающего протона, $mc^2$ - энергия протона мишени. Снова приравняем между собой значения импульсно-энергетического инварианта в лабораторной системе для in- состояния и системе центра инерции для out- состояния.

$$(E + mc^2)^2 - p^2c^2 = (4mc^2)^2 \qquad (5.11)$$

Раскроем скобки слева и учтем, что для протона $E^2 - p^2c^2 = m^2c^4$. Тогда окончательно будем иметь для полной и кинетической энергии «порогового» протона:

$$E = 7mc^2, \quad T = E - mc^2 = 6mc^2 \qquad (5.12).$$

Полученный результат показывает, что в лабораторной системе координат пороговая кинетическая энергия налетающей частицы ($6mc^2$) втрое превышает недостающую для протекания реакции внутреннюю энергию ($2mc^2$). Избыток в $4mc^2$ требуется для поддержания движения out- системы как целого (такое движение необходимо для выполнения закона сохранения импульса).

### 1.6. Сложение скоростей в релятивистской механике

До создания теории относительности явно или неявно, кинематика рассматривалась скорее не как раздел физики, а как раздел аналитической геометрии. Развитие релятивистской механики показало, что это не так. Оказалось, что законы кинематики являются вторичными по отношению к законам динамики в том смысле, что не могут быть априори навязаны физике, а устанавливаются апостериори в соответствии с принципами динамики материальных объектов. Поэтому вполне естественно, что с уточнением (усовершенствованием) законов динамики должны уточняться и законы кинематики. Проиллюстрируем эту мысль на основе релятивистского закона сложения скоростей.

Покажем, что существование релятивистского импульсно- энергетического инварианта приводит к релятивистскому закону сложения скоростей. Пусть в лабораторной системе координат рассматривается система из двух тел, движущихся вдоль одной и той же прямой со скоростями $v_1$ и $v_2$ соответственно. Попробуем определить, что увидит наблюдатель, движущийся вместе с телом №1 (например, барон Мюнхгаузен на ядре).

В лабораторной системе координат имеем для энергии и импульса частиц

$$E_1 = \frac{m_1 c^2}{\sqrt{1 - \frac{v_1^2}{c^2}}}, \quad E_2 = \frac{m_2 c^2}{\sqrt{1 - \frac{v_2^2}{c^2}}} \qquad (6.1)$$

$$p_1 = \frac{m \cdot v_1}{\sqrt{1 - \frac{v_1^2}{c^2}}}, \quad p_2 = \frac{m \cdot v_2}{\sqrt{1 - \frac{v_2^2}{c^2}}} \qquad (6.2)$$

В системе Мюнхгаузена тело №1 покоится, поэтому

$$p_1' = 0 \qquad E_1' = m_1 c^2, \qquad (6.3)$$

$$E'_2 = \frac{m_2 c^2}{\sqrt{1 - \dfrac{{v'_2}^2}{c^2}}} \qquad (6.4)$$

Из соотношения

$$\left(E_1 + E_2\right)^2 - c^2\left(\vec{p}_1 + \vec{p}_2\right)^2 = \left(E'_1 + E'_2\right)^2 - c^2\left(\vec{p}'_1 + \vec{p}'_2\right)^2 = \\ = E_0^2 = const \qquad (6.5)$$

путем несложных алгебраических выкладок получим тождество:

$$E_1 E_2 - c^2 p_1 p_2 = E'_1 E'_2 = const \qquad (6.6).$$

Из последнего выражения можно получить, что:

$${v'_2}^2 = \frac{(v_2 - v_1)^2}{\left(1 - \dfrac{v_1 v_2}{c^2}\right)^2}, \qquad (6.7)$$

откуда окончательно имеем:

$$v'_2 = \frac{v_2 - v_1}{1 - \dfrac{v_1 v_2}{c^2}} \qquad (6.8)$$

Заметим, что результат последнего преобразования, очевидно, определен с точностью до знака. Изменение знака в (6.8) отвечало бы только тому, что направление координатной оси в «движущейся» и «покоящейся» системах выбрано не параллельно а анти- параллельно.

Очевидно, что если частицы движутся навстречу друг другу, то вместо (6.8) имеем:

$$v'_2 = \frac{v_2 + v_1}{1 + \dfrac{v_1 v_2}{c^2}} \qquad (6.9)$$

Релятивистский закон сложения скоростей (6.8)- (6.9) снова подтверждает инвариантный и предельный характер скорости света (Мюнхгаузен, направляющийся «вдогонку» за квантом света, будет регистрировать ту же самую скорость света $c$, что и его более «благоразумный» коллега в покоящейся системе координат).

В нерелятивистском приближение, очевидно, из (6.8)- (6.9) следует классический (галилеевский) закон сложения скоростей.

Предоставляем читателю в качестве упражнения показать, что рассуждения, аналогичные проведенным выше, но с использованием классического импульсно- энергетического инварианта, непосредственно ведут к нерелятивистскому закону сложения скоростей.

Нередко при изучении релятивистской механики у читателя может создаться впечатление о предмете как о некоторой фантастической теории, для проверки которой, чтобы, скажем, убедиться в справедливости релятивистского закона сложения скоростей, обнаружить «парадокс близнецов», лоренцево сокращение и т.п., нужны конструкции типа «фотонных» ракет и пр. Особенно часто такое впечатление возникает из книг, нацеленных на популярное изложении. На самом деле, конечно, для того, чтобы убедится в справедливости релятивистской механики, нет необходимости выходить куда- либо за пределы физической лаборатории. Вместо проведенного выше описания с участием Мюнхгаузена, в реальной задаче можно говорить, например, об описании электрон- протонного рассеяния. Закон сложения скоростей в этом случае просто отвечает на вопрос о том, как перейти от описания рассеяния движущегося электрона на движущемся протоне к некоторому другому эквивалентному эксперименту, в котором, скажем, движущийся электрон рассеивается на покоящемся протоне. Другими словами, результаты двух различных экспериментов по электрон-протонному рассеянию будут изоморфны друг другу, если скорости сталкивающихся частиц, как параметры этих экспериментов, будут удовлетворять релятивистскому закону сложения скоростей (для перехода от параметров одного эксперимента к параметрам другого необходимо также использовать релятивистские формулы для преобразования и других величин, в том числе энергий, импульсов, поляризаций, углов, координат и др.).

### 1.7. В погоне за квантом (эффект Доплера)

Посмотрим, к чему приведет формула (6.6) из предыдущего параграфа, если частица №2– фотон.

$$\frac{m_1 c^2}{\sqrt{1-\frac{v_1^2}{c^2}}} \hbar\omega - c^2 \frac{m_1 v_1}{\sqrt{1-\frac{v_1^2}{c^2}}} \frac{\hbar\omega}{c} = m_1 c^2 \cdot \hbar\omega' , \qquad (7.1)$$

откуда:
$$\omega' = \frac{\omega(1-\beta)}{\sqrt{1-\beta^2}} , \qquad (7.2)$$

где $\beta = \frac{v_1}{c}$ - отношение скорости Мюнхгаузена к скорости света.

Здесь $\omega$ - частота, которую измерит наблюдатель в лабораторной системе координат, $\omega'$ - частота, которую будет наблюдать Мюнхгаузен.

### 1.8. Преобразования Лоренца

С использованием полученного выше релятивистского закона сложения скоростей нетрудно получить известные преобразования Лоренца, лежащие в основе релятивистской механики.

Рассмотрим движение тех же самых, что и в разд. 1.6 частиц, считая, что в нулевой момент времени они начинают двигаться из начала координат. Законы движения тел   с точки зрения наблюдателя в лабораторной системе координат, очевидно, есть:

$$x_1 = v_1 \cdot t_1 \qquad x_2 = v_2 \cdot t_2 \qquad (8.1)$$

С точки зрения Мюнхгаузена законы движения тех же тел есть:

$$x_1' = 0 \qquad x_2' = v_2' \cdot t_2' \qquad (8.2)$$

Согласно закону сложения скоростей имеем:

$$\frac{x_2'}{t_2'} = \frac{\left(\frac{x_2}{t_2} - v_1\right)}{1 - \frac{v_1 x_2}{c^2 t_2}} = \frac{x_2 - v_1 t_2}{t_2 - \frac{v_1 x_2}{c^2}} \qquad (8.3)$$

Для того, чтобы это соотношение выполнялось для всех тел в любые моменты времени, нужно, чтобы с точностью до некоторого множителя $\gamma$ числитель и знаменатель слева совпадал соответственно с числителем и знаменателем справа, т.е.

$$x' = \gamma(x - v \cdot t)$$
$$t' = \gamma\left(t - \frac{v \cdot x}{c^2}\right) \qquad (8.4)$$

Здесь мы несколько изменили исходные обозначения, убрав уже ненужные индексы. Теперь $x, t$ - это координата и время, отвечающие некоторому событию в лабораторной системе координат, $x', t'$ - координата и время, отвечающие тому же событию в системе Мюнхгаузена, скорость которого мы обозначили как $v$. Неизвестный коэффициент $\gamma$ нетрудно найти из требования эквивалентности прямых и обратных преобразований. Обратные к (8.4) преобразования координат должны, очевидно, отличаться только знаком скорости, т.е.

$$x = \gamma(x' + v \cdot t')$$
$$t = \gamma\left(t' + \frac{v \cdot x'}{c^2}\right) \qquad (8.5)$$

Сопоставляя (8.4) и (8.5), имеем

$$x = \gamma(x' + v \cdot t') = \gamma^2\left(x - vt + vt - \frac{v^2 \cdot x}{c^2}\right) = \gamma^2(1 - \beta^2)x, \text{ где } \beta = \frac{v}{c}$$

Отсюда получаем для релятивистского множителя:

$$\gamma = \frac{1}{\sqrt{1 - \frac{v^2}{c^2}}} \qquad (8.6)$$

Окончательно преобразования Лоренца запишем в виде:

$$x' = \frac{x - v \cdot t}{\sqrt{1 - \frac{v^2}{c^2}}}, \quad y' = y, \quad z' = z, \quad t' = \frac{t - \frac{v}{c^2}x}{\sqrt{1 - \frac{v^2}{c^2}}} \qquad (8.7)$$

Здесь мы учли также, что координаты $y$ и $z$ не меняются в рассматриваемом преобразовании.

Мы не будем рассматривать элементарные следствия преобразований Лоренца такие, как лоренцевское сокращение длин или замедление времени. Изложение этих вопросов читатель найдет в курсах общей физики.

Отметим также, что рассмотренное здесь преобразование, очевидно, является лишь частным случаем релятивистских преобразований координат. Оно называется буст- преобразованием Лоренца. Более общее преобразование координат может отличаться от описанного произвольным направлением относительной скорости, поворотом координатных осей, а также сдвигом начала отсчета координат и времени.

Нашей целью было проследить физические основы зарождения релятивистской механики в духе ее соответствия с механикой классической. Дальнейшее развитие релятивистских представлений, связанных с группой Пуанкаре, четырехмерным псевдо- евклидовым пространством Минковского, их приложением в электродинамике, теории калибровочных полей, гравитации и др. читатель может найти в современных учебниках по теоретической физике и в монографиях.

## Часть 2. Квантовая механика как рациональное статистическое обобщение классической механики (корневое статистическое квантование)

*Я полагаю, что понятие амплитуды вероятности, по-видимому, является наиболее фундаментальным понятием квантовой теории.*

Поль Дирак ([17], с.148)

### 2.1. Статистическая форма основного закона динамики

В процедуре перехода от классической механики к квантовой решающую роль играют статистические аргументы. Покажем, что квантовая механика является рациональным статистическим обобщением классической механики.

Рассмотрим самый обычный «школьный» второй закон Ньютона:

$$\frac{d^2}{dt^2}\vec{x} = -\frac{1}{m}\frac{\partial U}{\partial \vec{x}} \tag{1.1}$$

Предположим, что фигурирующие в этом законе ускорение и сила есть некоторые средние величины. Усреднение обеспечивается посредством введения некоторой плотности распределения $P(x)$:

$$\frac{d^2}{dt^2}\left(\int P(x)\vec{x}\,dx\right) = -\frac{1}{m}\left(\int P(x)\frac{\partial U}{\partial \vec{x}}\,dx\right) \tag{1.2}$$

Таким образом, мы отказываемся от детерминистской формы основного закона динамики. В силу принципа соответствия, однако, мы хотим потребовать, чтобы законы классической динамики оставались справедливыми в среднем. Формула (1.2), очевидно, является более общей по сравнению с (1.1) и включает последнюю в качестве частного (предельного) случая, отвечающего дельта- образной плотности распределения.

Пусть плотность распределения $P(x)$ задана в виде многопараметрической зависимости $P(x|c_1(t), c_2(t), ..., c_s(t))$. Динамика плотности в этом случае определяется изменением параметров $c_1(t), c_2(t), ..., c_s(t)$ во времени. Заметим, что такой (параметрический) подход с самого начала не опирается ни на какие представления о траекториях частиц. Да, «облако», описываемое плотностью $P(x)$, меняется со временем, но это обусловлено просто тем, что меняются параметры $c_1(t), c_2(t), ..., c_s(t)$, однако за этими изменениями не стоят никакие траектории.

Обычно статистические закономерности связывают с хаотическим (стохастическим) движением частиц. В основе такого рассмотрения лежит представление о том, что у частиц существуют какие- то («скрытые») траектории, просто мы их не знаем, поскольку движение «сложное». Такое представление о случайности является субъективистским. Оно не имеет никакого отношения к описанию микроявлений и к квантовой механике в целом. Природа самих явлений не может зависеть от нашего знания или незнания. Как следует из классической механики и электродинамики, электрон, если бы он двигался по траектории, непрерывно излучал бы энергию и неизбежно очень быстро упал бы на ядро (и наше незнание его траектории, очевидно, никак не сделало бы атом устойчивым). Для того, чтобы адекватно представлять себе явления в квантовой механике, необходимо исходить из представления о вероятности как объективной категории. Мы пишем плотность $P(x)$ не потому, что мы не знаем «истинных» траекторий, а потому что и сама Природа их не знает (и даже не интересуется их существованием). Подчеркнем еще раз, что для статистического описания, задаваемого функцией плотности $P(x|c_1(t), c_2(t),...,c_s(t))$, нет никакой необходимости вводить какие- либо траектории, поскольку динамика плотности непосредственно индуцируется параметрами распределения $c_1(t), c_2(t),..., c_s(t)$.

### 2.2. Корневая модель и матричное уравнение Гейзенберга

Оказывается, что среди всевозможных многопараметрических представлений плотности существует одно определенное представление, которое выделяется среди других своими наиболее простыми и фундаментальными статистическими свойствами. Это так называемое **корневое** разложение [18,19]. Оно задает плотность распределения в виде $P(x)$:

$$P(x) = |\psi|^2, \tag{2.1}$$

где $\psi(x) = c_j \varphi_j(x)$ (2.2)

Здесь $\varphi_j(x)$ - ортонормированный базис. В формуле (2.2) предполагается суммирование по повторяющемуся индексу $j$.

Модель (2.1) мы называем корневой, поскольку она основана на введении нового объекта – пси функции, которая является как- бы квадратным корнем из плотности. В физической литературе связь между пси- функцией и плотностью называется формулой Борна. Подчеркнем, однако, что выделенный характер такого представления плотности по сравнению с любым другим следует уже из статистических соображений и никак не связан априори с квантовой механикой. Просто Природа «выбрала» эту модель как оптимальную среди всех возможных моделей статистического оценивания. Мы не можем здесь более подробно рассматривать этот вопрос и отсылаем читателя к литературе [20- 21]. Подчеркнем только, что с точки зрения корневого подхода пси функция математическим объектом для статистического описания данных. Корневое разложение находит применение в задачах квантовой томографии, связанных с восстановлением квантовых состояний по результатам взаимно- дополнительных измерений [22- 24].

Предположим, что зависимость коэффициентов разложения от времени соответствует гармоническим колебаниям:

$$c_j(t) = c_{j0} \exp(-i\omega_j t) \tag{2.3}$$

Подчеркнем, что собственные частоты колебаний механической системы и базисные функции разложения заранее неизвестны. Их следует определить таким образом, чтобы выполнялись усредненные уравнения движения. Покажем, что модель, задаваемая уравнением (1.2) совместно с условиями (2.1)- (2.3) приводит к стационарным функциям и частотам уравнения Шредингера.

Подставляя (2.1)-(2.3) в (1.2), получим:

$$m(\omega_j - \omega_k)^2 c_{j0} c_{k0}^* \langle k|\vec{x}|j\rangle \exp(-i(\omega_j - \omega_k)t) = $$
$$= c_{j0} c_{k0}^* \langle k|\frac{\partial U}{\partial \vec{x}}|j\rangle \exp(-i(\omega_j - \omega_k)t) \tag{2.4}$$

Здесь, как обычно, по повторяющимся индексам $j$ и $k$ предполагается суммирование.

Матричные элементы в выражении (2.4) определяются формулами:

$$\langle k|\vec{x}|j\rangle = \int \varphi_k^*(x)\vec{x}\,\varphi_j(x)dx \tag{2.5}$$

$$\langle k|\frac{\partial U}{\partial \vec{x}}|j\rangle = \int \varphi_k^*(x)\frac{\partial U}{\partial \vec{x}}\varphi_j(x)dx \tag{2.6}$$

Для того, чтобы соотношение (2.4) выполнялось в любой момент времени для произвольных начальных амплитуд, следует потребовать выполнения равенства левых и правых частей отдельно для каждого слагаемого в сумме, поэтому:

$$m(\omega_j - \omega_k)^2 \langle k|\vec{x}|j\rangle = \langle k|\frac{\partial U}{\partial \vec{x}}|j\rangle \tag{2.7}$$

Последнее выражение представляет собой матричное уравнение Гейзенберга для квантовой динамики в энергетическом представлении.

В работе [7] Бор пишет о матричной механике Гейзенберга как адекватном выражении принципа соответствия: «только благодаря квантово- теоретическим методам, созданным за последние несколько лет (*имеется ввиду период 1925- 1928 г.г. – Ю.Б.*), общие стремления, заложенные в упомянутом принципе (*соответствия*), получили адекватную формулировку». Интерпретация результатов вычислений в квантовой механике первое время после ее открытия сильно осложнялась из- за непонимания ее статистического характера. Статистическая интерпретация теории впервые дана Борном.

Матричное уравнение Гейзенберга задает условия, которым должны удовлетворять базисные функции и частоты механической системы, чтобы, в среднем, движение удовлетворяло основному закону динамики. Заметим что, этих условий очень много (вообще говоря, бесконечно много). Так, если мы хотим ограничиться приближением, в котором учитывается сто первых функций и частот, то нам следует наложить на них десять тысяч условий. Заранее не очевидно, что это вообще можно сделать. Решение этой задачи, однако, существует и фактически представляет собой переход от гейзенберговской картины к шредингеровской. Прежде чем явно сконструировать это решение, сделаем два замечания относительно уже полученных результатов.

Замечание 1. В силу линейности (по плотности) основного закона динамики в статистической форме (1.2), формула $P(x) = |\psi(x)|^2$ допускает очевидное обобщение (модель смеси из $s$ компонент):

$$P(x) = |\psi^{(1)}(x)|^2 + |\psi^{(2)}(x)|^2 + ... + |\psi^{(s)}(x)|^2 \tag{2.8}$$

К каждой компоненте смеси применимы все вышеизложенные соображения. Наличие нескольких компонент плотности вместо одной никак не меняет наших рассуждений, поскольку усредненный закон динамики выполняется для каждой из компонент независимо. Это простое обобщение позволяет формально рассмотреть не только «чистые» состояния, задаваемые (2.1) но и так называемые «смешанные» состояния, задаваемые (2.8).

Замечание 2. Любые другие модели, кроме $P(x) = |\psi(x)|^2$ (например $P(x) = \psi(x)$,

$P(x)=|\psi(x)|^3$, $P(x)=\ln(|\psi(x)|)$, $P(x)=\exp(-|\psi(x)|^2)$ и др.) не приводят к замкнутому условию на собственные частоты и базисные функции, аналогичному матричному уравнению Гейзенбергу, так как в сумме не удается избавиться от зависимости от времени и начальных амплитуд. Таким образом, только корневая модель для параметризации плотности приводит к самосогласованной задаче статистического обобщения классической механики.

### 2.3. Переход от гейзенберговской картины к шредингеровской

Покажем, что базисные функции и частоты, удовлетворяющие матричному соотношению (2.7), есть стационарные состояния и частоты квантовой системы (в соответствии с эквивалентностью картин Гейзенберга и Шредингера).

Действительно, образуем диагональную матрицу из частот системы $\omega_j$. Рассматриваемая матрица будет эрмитовой в силу того, что частоты – действительные числа. Эта матрица будет представлением некоторого эрмитова оператора, собственные значения которого суть $\omega_j$, т.е.

$$\hat{\Omega}|j\rangle = \omega_j |j\rangle, \tag{3.1}$$

Найдем явный вид искомого оператора частоты $\hat{\Omega}$. В силу (3.1), матричное соотношение (2.7) можно переписать в виде операторного уравнения

$$\left[\hat{\Omega},\left[\hat{\Omega},\vec{x}\right]\right] = \frac{1}{m}\vec{\partial}U, \tag{3.2}$$

где $\vec{\partial} = \dfrac{\partial}{\partial \vec{x}}$ - оператор дифференцирования, [ ] - коммутатор.

Выражение, стоящее в правой части (3.2), представим в виде некоторого коммутатора:

$$\frac{1}{m}\vec{\partial}U = \left[\frac{1}{\hbar}U, -\frac{\hbar}{m}\vec{\partial}\right], \tag{3.3}$$

где $\hbar$ – произвольная константа, которая, в итоге, должна быть отождествлена с постоянной Планка (см. обсуждение ниже).

Рассматриваемый коммутатор, очевидно, не изменится, если к потенциальной составляющей $\dfrac{1}{\hbar}U$ добавить произвольную функцию от оператора производной $F_1(\vec{\partial})$, т.е.

$$\frac{1}{m}\vec{\partial}U = \left[\frac{1}{\hbar}U, -\frac{\hbar}{m}\vec{\partial}\right] = \left[F_1(\vec{\partial}) + \frac{1}{\hbar}U, -\frac{\hbar}{m}\vec{\partial}\right] \tag{3.4}$$

Здесь в правой части остается еще произвол, связанный с возможностью добавления к оператору $-\dfrac{\hbar}{m}\vec{\partial}$ произвольного постоянного вектора. Этот произвол, однако, несуществен.

Аналогичным образом имеем:

$$-\frac{\hbar}{m}\vec{\partial} = \left[-\frac{\hbar}{2m}\partial^2, \vec{x}\right] = \left[-\frac{\hbar}{2m}\partial^2 + F_2(\vec{x}), \vec{x}\right], \tag{3.5}$$

где $F_2(\vec{x})$ - произвольная функция от координат.

Таким образом:

$$\left[\hat{\Omega}[\hat{\Omega}, \vec{x}]\right] = \left[F_1(\vec{\partial}) + \frac{1}{\hbar}U, \left[-\frac{\hbar}{2m}\partial^2 + F_2(\vec{x}), \vec{x}\right]\right] \tag{3.6}$$

Последнее соотношение оказывается согласованным, если положить:

$$F_1(\vec{\partial}) = -\frac{\hbar}{2m}\partial^2 \qquad F_2(\vec{x}) = \frac{1}{\hbar}U \tag{3.7}$$

Окончательно находим, что решением уравнения (3.2) является оператор:

$$\hat{\Omega} = -\frac{\hbar}{2m}\partial^2 + \frac{1}{\hbar}U(x) \tag{3.8}$$

Для того, чтобы слагаемые в (3.8) имели одинаковую размерность, произвольная константа $\hbar$ должна иметь размерность постоянной Планка (эрг*с). Численное значение этой постоянной должно быть выбрано таким, чтобы собственные значения оператора частоты $\hat{\Omega}$ совпадали с реальными атомными частотами. Нетрудно видеть, что выбор численного значения постоянной Планка $\hbar$ связан с выбором единиц измерения для основных физических величин (длина, время, масса). С теоретической точки зрения единицы измерений можно выбрать так, чтобы было $\hbar = 1$ (заметим, что в квантовой теории поля общеупотребительна система единиц, в которой $\hbar = c = 1$).

Вместо оператора частоты $\hat{\Omega}$ в квантовой теории принято использовать гамильтониан $\hat{H}$.

$$\hat{H} = \hbar\hat{\Omega} = -\frac{\hbar^2}{2m}\partial^2 + U(x) \tag{3.9}$$

Собственные значения гамильтониана согласно (3.1) есть:

$$\hat{H}|j\rangle = \hbar\omega_j|j\rangle \tag{3.10}$$

Таким образом, если потребовать, чтобы корневая оценка плотности удовлетворяла в среднем классическим уравнениям движения, то базисные функции и частоты корневого разложения уже не могут быть произвольными, а должны представлять собой соответственно собственные функции и собственные значения гамильтониана системы.

Рассмотрим теперь модель смеси и построим для нее матрицу плотности, элементы которой определим формулой (сумму по компонентам смеси выписываем явно):

$$\rho_{jk} = \sum_{l=1}^{s} c_j^{(l)} c_k^{*(l)} = \sum_{l=1}^{s} c_{j0}^{(l)} c_{k0}^{*(l)} \exp(-i(\omega_j - \omega_k)t) \tag{3.11}$$

На основе представленных выше результатов нетрудно получить уравнение для динамики матрицы плотности, называемое обычно квантовым уравнением Лиувилля:

$$\frac{\partial \hat{\rho}}{\partial t} = -\frac{i}{\hbar}\left[\hat{H}, \hat{\rho}\right] \tag{3.12}$$

**2.4 Оператор импульса**

С использованием полученного выражения (3.9) для гамильтониана уже нетрудно получить операторные представления для других динамических величин. Например, понятие импульса можно ввести на основе следующей легко проверяемой цепочки равенств:

$$m\frac{d}{dt}\left(\int P(x)\vec{x}dx\right) = -im(\omega_j - \omega_k)c_{j0}c_{k0}^*\langle k|\vec{x}|j\rangle \exp(-i(\omega_j - \omega_k)t) =$$
$$= \frac{im}{\hbar}\langle \psi|Hx - xH|\psi\rangle = \langle \psi|\hat{\vec{p}}|\psi\rangle \tag{4.1}$$

В выражении (4.1) суммирование по индексам $j$ и $k$ предполагается автоматически.

Первое из представленных равенств непосредственно следует из определения корневого разложения плотности (2.1)- (2.2), при получении второго равенства мы учли (3.10), наконец последнее равенство следует из определения импульса (в нерелятивистской теории оператор импульса должен быть определен таким образом, чтобы его среднее значение совпадало с произведением массы на среднюю скорость).

Из соотношения (4.1) с необходимостью вытекает следующее определение импульса:

$$\hat{\vec{p}} = \frac{im}{\hbar}\left[\hat{H}\vec{x}\right] = -i\hbar\frac{\partial}{\partial \vec{x}} \tag{4.2}$$

Заметим, что выражения для операторов наблюдаемых величин мы не постулируем (как это делают при стандартном изложении квантовой механики), а <u>выводим</u> как необходимые следствия корневого разложения плотности.

Зная вид оператора импульса, легко получаем коммутационное соотношение между его компонентами и координатами:

$$\left[p_j, x_k\right] = -i\hbar\delta_{jk} \tag{4.3}$$

Дираком была показана аналогия между коммутационными соотношениями (4.3) и фундаментальными классическими скобками Пуассона. На этой аналогии может быть построена теория квантовых канонических преобразований.

В целом, далее при изложении квантовой механики можно следовать стандартным, хорошо зарекомендовавшим себя курсам.

Не секрет, что математический аппарат квантовой механики очень сильно отличается от формализма классической механики. В результате, при изучении квантовой механики очень часто в сознании слушателей возникает «брешь», связанная с предметным (физическим) непониманием, как самой квантовой теории, так и ее связей с классической теорией. На наш взгляд, указанная «брешь» может быть ликвидирована, только если рассматривать квантовую теорию как такую теорию, в самой основе которой лежат статистические законы. Мы с самого начала ввели в рассмотрение пси функцию как математический объект статистического описания. При таком изложении квантовой механики статистический характер теории – это не интерпретация, а сама ее сущность.

Соотношения, согласно которым, уравнения классической механики выполняются в среднем и для квантовых систем, называют уравнениями Эренфеста

[25]. Самих этих уравнений, конечно, недостаточно для описания квантовой динамики. Как было показано выше, дополнительное условие, которое позволяет преобразовать классическую механику в квантовую (т.е. условие квантования), есть, по- существу, требование корневого характера плотности.

**Заключение**

Сформулируем основные результаты работы

1. Принцип соответствия Н. Бора показывает, что релятивистская механика, ровно как и квантовая механика могут рассматриваться как рациональные обобщения классической механики. Обе «новые» механики содержат «старую» в качестве частного (предельного) случая. Результаты релятивистской механики переходят в результаты классической механики, когда рассматривается движение тел со скоростями, малыми по сравнению со скоростью света в вакууме. Аналогично, результаты квантовой механики согласуются с результатами классической механики, когда характерные для задачи параметры размерности действия велики по сравнению с постоянной Планка.
2. Проведено сравнительное изложение классической и релятивистской механик с использованием трех основных положений: определения импульса, основного закона динамики (второго закона Ньютона) и закона сохранения энергии. Отличие релятивистской механики от классической обусловлено новым определением импульса как меры количества движения, пропорциональной скорости и энергии.
3. Рассмотрена связь энергии с импульсом в нерелятивистской и релятивистской механике. Получены импульсно- энергетические инварианты для классической и релятивистской систем. Эти инварианты позволяют связывать между собой характеристики in- состояний (до взаимодействия) и out- состояний (после взаимодействия) в различных системах отсчета.
4. Показано, что новые релятивистские законы динамики приводят к необходимости видоизменения кинематических соотношений классической механики, таких как закон сложения скоростей, преобразование координат, частот и др.
5. Показано, что квантовая механика может рассматриваться как рациональное статистическое обобщение классической механики. Учет принципа соответствия обеспечивается посредством требования, чтобы новое статистическое описание согласовывалось в среднем с классическим основным законом динамики. Статистические закономерности в квантовой механике носят фундаментальный объективный характер и не связаны с неполнотой информации об изучаемой системе.
6. Среди возможных многопараметрических статистических моделей выделенную роль играет корневая модель, связанная с введением амплитуд вероятностей (пси функции) как математического объекта статистического анализа данных. Построение многопараметрической статистической модели сводится к нахождению таких частот и базисных функций в разложении Фурье, которые обеспечивали бы выполнение в среднем уравнений движения. Только корневая модель приводит к согласованному условию, связывающему собственные частоты и функции механической системы и выражаемому матричным уравнением Гейзенберга.
7. Матричное уравнение Гейзенберга сводится к операторному уравнению, решение которого можно интерпретировать как построение гамильтониана системы и переход к картине Шредингера.

8. Рассматриваемый подход естественным образом приводит к понятию оператора импульса, фундаментальным коммутационным соотношениям, построению матрицы плотности, уравнения Лиувилля и др.

**Список литературы**